\documentclass[11pt]{amsart}

\usepackage{hyperref}

\usepackage{graphicx, enumerate, url}
\usepackage{amssymb,mathtools}
\usepackage{algorithm}
\usepackage{algpseudocode}
\usepackage{color}
\usepackage{tikz}
\usetikzlibrary{3d,calc}

\numberwithin{equation}{section}
\numberwithin{algorithm}{section}

\theoremstyle{plain}
\newtheorem{theorem}{Theorem}[section]

\newtheorem{lemma}[theorem]{Lemma}

\theoremstyle{definition}

\theoremstyle{remark}
\newtheorem{remark}[theorem]{Remark}

\let\ol\overline
\def\df{\overset{\mathrm{df}}{=}}
\def\bbI{\mathbb{I}}

\newcommand{\haf}{\mathop{{\mathrm{haf}}}\nolimits}
\newcommand{\per}{\mathop{{\mathrm{per}}}\nolimits}

\newcommand{\floor}[1]{\lfloor #1\rfloor}

\newcommand{\cM}{\mathcal{M}}

\newcommand{\rS}{\mathrm{S}}

\begin{document}
\title{Nonnegativity for hafnians of certain matrices}
\author[Kamil Br\'adler]{Kamil~Br\'adler}
\address{ORCA Computing. Formerly Xanadu.}
\email{kamilbradler@gmail.com}
\author[Shmuel Friedland]{Shmuel~Friedland}
\address{Department of Mathematics and Computer Science, University of Illinois at Chicago, Chicago, Illinois, 60607-7045, USA }
\email{friedlan@uic.edu}
\author[Robert B. Israel]{Robert~B.~Israel}
\address{Department of Mathematics, University of British Columbia, Vancouver, BC, Canada, V6T 1Z2}
\email{israel@math.ubc.ca}
\begin{abstract}
We show that a complex symmetric matrix of the form
$A(Y,B)=\begin{bmatrix}
         Y & B \\
         B^\top & \ol{Y} \\
\end{bmatrix}$,
where $B$ is Hermitian positive semidefinite, has a nonnegative hafnian.   Some positive scalar multiples of  matrices $A(Y,B)$ are encodable in a Gaussian boson sampler.  Further, the hafnian of this matrix is non-decreasing in $B$ in the sense that $\haf{A}(Y,L) \ge \haf{A}(Y,B)$ if $L \succeq B$.
\end{abstract}
\maketitle
 \noindent {\bf 2020 Mathematics Subject Classification.} 15A15, 15B57, 15A45, 81V73.
\noindent \emph{Keywords}:  Hafnians, permanents, positive semidefnite hermitian matrices.
\maketitle

\section{Introduction} \label{sec:intro}

Let $A = [a_{ij}]$ be a $2n \times 2n$ symmetric matrix with entries in $\mathbb C$.  The hafnian $\haf A$ is defined as the sum of $\prod_{k=1}^n a_{i_k j_k}$ over all
perfect matchings $(i_1,j_1),\ldots,(i_n,j_n)$ of the complete graph $K_{2n}$.
The pairs $(i,j)$ for which $i \ne j$ and $a_{ij} \ne 0$ form the edges of a graph $G$ with vertex set $[2n]$; we
can consider $A$ (with diagonal entries ignored) as a weighted adjacency matrix of $G$, and $\haf A$ is
a weighted sum over the perfect matchings of $G$.


Assume that $m=2n$ and $K_{2n}=([2n],E_{2n})$ is a complete graph on $[2n]$ vertices.  Recall that $M\subset E_{2n}$ is a perfect match of $K_{2n}$ if $([2n],M)$ is a $1$-regular spanning subgraph of $K_{2n}$.  So $M=\bigcup_{k\in[n]}\{(i_k,j_k)\}$, where $[2n]=\bigcup_{k\in[n]}\{i_k,j_k\}$.  Let $\cM_{2n}$ be the set of perfect matchings in $K_{2n}$.
Assume that $A=[a_{ij}]\in\rS_{2n}$.

Then the hafnian of $A$ is defined as follows \cite{caianiello1953quantum}:
\begin{eqnarray*}
\haf A=\sum_{M=\bigcup_{k\in[k]}(i_k,j_k)\in\cM_{2n}}\prod_{k=1}^n a_{i_kj_k}.
\end{eqnarray*}
For various properties of the hafnian see, e.~g.,~\cite{barvinok2017combinatorics}.

In particular we consider $A$ of the form
\begin{align}\label{eq:bpform}
A(Y,B)=\begin{bmatrix}
         Y & B \\
         B^\top & \ol{Y} \\
       \end{bmatrix},
\end{align}
where $Y$ is a (complex) symmetric matrix and $B$ is a hermitian positive semidefinite matrix.  The main result of this paper is
%
\begin{theorem}\label{thm:mainthm}
Assume that $A(Y,B)$ is of the form \eqref{eq:bpform}, where $Y$ is complex symmetric and $B$ positive semidefinite Hermitian.  Then $\haf{A(Y,B)}\ge 0$.
If $B$ has no zero row then $\haf{A(Y,B)}> 0$.
  Furthermore, if $L\succeq B$ i.e., $L-B$ is positive semidefinite Hermitian then $\haf{A(Y,L)}\ge \haf{A(Y,B)}$.
\end{theorem}

The inequality $\haf{A(Y,B)}\ge 0$ for $B\succeq 0$ can be deduced from the physical arguments stated in \cite{bradler2018graph}.  See Appendix.

Observe that $A(0,B)$ for any $B \in \mathbb C^{n \times n}$ is a weighted adjacency matrix of the complete bipartite graph $K_{n,n}$, where the first part is $[n]$ and the second part is $[n+1,\ldots,2n]$. Permutations of $[n]$ correspond to perfect matchings of $K_{n,n}$, so that the permutation $\sigma$ corresponds to the matching
consisting of pairs $(j,n+\sigma(j))$.  Hence
%
\begin{equation*}
\haf A(0,B)=\per B=\sum_{\sigma\in\Sigma_n} \prod_{k=1}^n b_{k\sigma(k)}.
\end{equation*}
Assume that $B$ is positive semidefinite.  It is well known that $\per{B}\ge 0$. This is a corollary of Schur's theorem \cite{schur1918endliche} that for a positive semidefinite $B$ we have the inequality $\per B\ge \det B$. The latter is nonnegative (being the product of the eigenvalues of $B$).   See also \cite{bapat1986extremal}.
Moreover,  $\per B=0$ if and only if $B$ has a zero row \cite[Theorem 3]{MN62}.  Theorem \ref{thm:mainthm} yields that $\per L\ge \per B$ if $L\succeq B\succeq 0$.  This inequality  may be known but we did not find it in the literature.

\begin{remark}
  Note that the problem of computing the sign of the permanent is in general hard~\cite{aaronson2011linear}. Hence a similar result holds for the hafnian.
\end{remark}
\section{Proof of the main theorem}
Let $Q = (q_{st})$ be an $m\times  n$ complex valued matrix and denote the transpose and the conjugate transpose as $Q^\top$ and $Q^*$, respectively. The $r$-th induced matrix $P_r(Q)$ is defined as follows \cite[p. 20]{marcus1992survey}.  Denote by $G_{k,n}$ the totality of nondecreasing sequences of $k$ integers chosen from $[n]=\{1,\ldots,n\}$.  Let $\alpha\in G_{k,n}$. Then $\mu(\alpha)$ is defined to be the product of the factorials of the multiplicities of the distinct integers appearing in the sequence $\alpha$. For $\alpha\in G_{k,m}, \beta\in G_{l,n}$ we set $Q[\alpha,\beta]\df(q_{\alpha_s\beta_t})_{\genfrac{}{}{0pt}{2}{s=1\dots k}{t=1\dots l}}$ to be the $k\times l$ submatrix of $Q$ with the rows and columns in $\alpha$ and $\beta$, respectively. Now $P_r(Q)$ is the ${m+r-1\choose r}\times {n+r-1\choose r}$ matrix whose entries are $ \per{Q[\alpha,\beta]}/ \sqrt{\mu(\alpha)\mu(\beta)}$ arranged lexicographically in $\alpha=(\alpha_1,\ldots,\alpha_r)\in G_{r,m}, \beta=(\beta_1,\ldots,\beta_r)\in G_{r,n}$.  Recall that $P_r(Q^*)=P_r(Q)^*$ and if $S$ is an $n\times p$ matrix then $P_r(QS)=P_r(Q)P_r(S)$ \cite{marcus1992survey}.

%

Assume that $B$ is an $m\times m$ Hermitian matrix. Then the spectral decomposition of $B$ is $UDU^*$ where $D$ is a real diagonal matrix.  Then
\[
P_r(UDU^*)=P_r(U)P_r(DU^*)=P_r(U)P_r(D)P_r(U^*)=P_r(U)P_r(D)P_r(U)^*.
\]
Clearly, if $D$ is a real diagonal matrix then $P_r(D)$ is also a diagonal matrix with real entries.  Hence $P_r(B)$ is Hermitian.  Assume that $B$ is positive semidefinite.  Hence $D$ is a nonnegative diagonal matrix.  It is straightforward to show that $P_r(D)$ is also a nonnegative diagonal matrix.  Hence, if $B$ is an $m\times m$ positive semidefinite Hermitian matrix then $P_r(B)$ is positive semidefinite.  Let $H$ be a diagonal matrix of order
${m+r-1\choose r}$ whose diagonal entries are $\sqrt{\mu(\alpha)}$.  If $B$ is positive semidefinite then the matrix $C_r(B)=HP_r(B)H$ is also positive semidefinite.  Note that the entries of $C_r(B)$ are $\per B[\alpha,\beta]$.

We now consider the hafnian of $A = A(Y,B)$, where $Y$ is complex symmetric and $B$ Hermitian.
A perfect matching of $[2n]$ will match some $\alpha \subseteq [n]$ with itself, while a subset $n + \beta, \beta\subset [n]$ of $n+[n] = \{n+1,\ldots,2n\}$ of equal cardinality is matched to itself,
and the remaining members $[n]\setminus \alpha$ of $[n]$ are matched to $n+([n] \setminus \beta)$.
The contribution to $\haf A(Y,B)$ of such matchings for a particular $\alpha$ and $\beta$ is
\[
    \haf (Y[\alpha,\alpha]) \per(B([n]\setminus \alpha,[n]\setminus \beta))\,\ol{\haf (Y[\beta,\beta])},
\]
where we take the hafnian or permanent of an empty matrix to be $1$.
The total contribution of all of these for a given $k$, $0 \le k \le \floor{n/2}$, is

\begin{align}\label{eq:cont2khaf}
 \sum_{\alpha: |\alpha|=2k} \sum_{\beta: |\beta|=2k}  \haf (Y[\alpha,\alpha]) \per(B([n]\setminus \alpha,[n]\setminus \beta))\,\ol{\haf (Y[\beta,\beta])}.
\end{align}

Note that the matrix $F_{n-2k}(B)$ whose entries are  $\per{B[\gamma,\delta]}$ for $\gamma,\delta$ all $n-2k$-subsets of $[n]$ is a principal submatrix of $C_{n-2k}(B)$, hence Hermitian,
and positive semidefinite if $B\succeq 0$.  Hence the sum~\eqref{eq:cont2khaf} is real and nonnegative if $B\succeq 0$.  This shows that $\haf{A(Y,B)} \ge 0$. Recall \cite[Theorem 3]{MN62} that $\per B>0$ if $B$ has no zero row. Hence $\haf A(Y,B)>0$ if $B$ has now zero row.

Assume now that $L\succeq B\succeq 0$.  We claim that $P_r(L)\succeq P_r(B)\succeq 0$.  (The last inequality was established above.)
Assume first that $\det B>0$, i.e., $B$ is positive definite.  Then $B$ has a unique positive definite square root $R$, and $L \succeq B$ is equivalent to $ L_1 \df R^{-1} L R^{-1} \succeq \bbI_n$,
where $\bbI_n$ is the identity matrix of order $n$. Thus we can diagonalize $L_1 = U D U^*$, where $U$ is unitary and $D$ is diagonal with diagonal entries and the eigenvalues of $L_1$ are all $\ge 1$.
  Recall that $P_r(\bbI_n)=\bbI_{n+r-1\choose r}$ \cite[2.12.5]{marcus1992survey}. Thus
\begin{align*}
  P_r(L_1)&=P_r(UDU^*)=P_r(U)P_r(D)P_r(U)^*, \\
  \bbI_{n+r-1\choose r}&=P_r(\bbI_n)=P_r(UU^*)=P_r(U) P_r(U)^*.
\end{align*}
As each diagonal entry of $D$ is at least $1$ we deduce that $P_r(D)\succeq P_r(\bbI_m)=\bbI_{n+r-1\choose r}$.  Thus, each eigenvalue of $P_r(L_1)$ is at least $1$.  Hence $P_r(L_1)\succeq\bbI_{n+r-1\choose r}$.  Observe next
\[
P_r(L_1)=P_r(R^{-1}LR^{-1})=P_r(R)^{-1}P_r(L)P_r(R)^{-1}\succeq \bbI_{n+r-1\choose r}.
\]
Use the previous observation to deduce that
$P_r(L)\succeq P_r(R)P_r(R)=P_r(R^2)=P_r(B)$.  This concludes the proof in the case that $B$ is nonsingular.  For the general case, we  note that $\haf A(Y, L+\epsilon \bbI_n) \ge \haf A(Y, B+\epsilon \bbI_n)$ for $\epsilon > 0$ and take the limit as $\epsilon \to 0^+$ (the hafnian being a continuous function).

\bibliographystyle{plain}

 \appendix
 \section{Gaussian Boson Sampling}\label{sec:GuasBS}
 In this appendix we discuss the connection of our results to Gaussian Boson Sampling.  To keep this section aligned with the notation of the physics literature we replace the integer $n$ by the integer $M$.

A link between hafnians of certain matrices and covariance matrices of quantum-optical Gaussian states was put forward in \cite{hamilton2016gaussian} and further explored in \cite{bradler2018graph}. Ref.~\cite{hamilton2016gaussian} introduced  a Gaussian boson sampler (GBS) as a generalization of the boson sampler~\cite{aaronson2011computational} where an $M$-mode linear interferometer is fed by a product of $M$ single-mode squeezed states  and its output is sampled by an array of $M$ photon number-resolving detectors. It turns out that the probability of detecting exactly one photon in each output detector is proportional to the hafnian of a certain matrix~$A$ (for a generalization to all possible multiphoton events see~\cite{bradler2018graph}).

The complex covariance matrix describing the input to the interferometer has dimension $2M\times2M$ and encodes the covariances of the canonical  operators ${\boldsymbol{\xi}}=(a_1,\dots,a_M$, $a_1^\dagger,\dots,a_M^\dagger$):
\begin{equation}\label{eq:covmatrix}
 \sigma_{ij} = \tfrac{1}{2}\langle{\xi}_i {\xi}_j + {\xi}_j {\xi}_i\rangle - \langle {\xi_i} \rangle\langle {\xi_j} \rangle.
\end{equation}
The symbol $\dagger$ denotes Hermitian conjugation and $\langle.\rangle$ denotes the operator expectation value. The physical covariance matrix is Hermitian, positive semidefinite and its symplectic eigenvalues are greater than $1/2$~\cite{weedbrook2012gaussian}.
The authors of~\cite{hamilton2016gaussian} did not offer the most general form of~$A$ leading to a physical covariance matrix. Instead, they use $A=\begin{bmatrix}
         Y & B \\
         B^\top & \ol{Y} \\
\end{bmatrix}$ for an arbitrary complex $B$ and complex symmetric~$Y$; however the corresponding covariance matrix may be non-physical. The physical relevance of knowing what $A$ can be encoded in the GBS device is related to the question of which weighted undirected graphs can have their hafnians sampled by a GBS device~\cite{bradler2017gaussian}. In~\cite{hamilton2016gaussian}  the canonical form $A=Y\oplus\ol{Y}$ was used, as this always leads  to a physical covariance matrix.  However, this comes at the expense of `doubling' the adjacency matrix~\cite{bradler2017gaussian}, leading to lower detection probabilities.

 We claim that Corollary~3 of~\cite{bradler2018graph} holds for complex matrices as well:
\begin{lemma}\label{lem:cor3} Let
$R=\begin{bmatrix}
    R_{11}&R_{12}\\
    R_{21}&R_{22}
\end{bmatrix}$
be a $2M\times2M$ complex symmetric matrix.  Then there exists a Gaussian covariance matrix $\sigma$ such that
\begin{equation}\label{eq:cRform}
cR=X_{2M}[\bbI_{2M}-(\sigma+\frac{1}{2}\bbI_{2M})^{-1}],
\end{equation}
where
\[
X_{2M}=\begin{bmatrix}
           0 & \bbI_M \\
           \bbI_M & 0 \\
         \end{bmatrix}
\]
if and only if:
\begin{enumerate}
\item $R_{11} = \ol{R}_{22}$ and $R_{12} = R_{21}^\top$.
\item $R_{12}$ is Hermitian and positive semidefinite.
\item $c\in (0, 1/\|R\|_2)$
\end{enumerate}
\end{lemma}
\begin{proof}  Since $R$ is complex symmetric we must have that $R_{11},R_{22}$ are complex symmetric and $R_{21}=R_{12}^\top$.  Set $Y=R_{11}, B=R_{12}$ in $\eqref{eq:bpform}$.   Let $F=cX_{2M}R$ for some $c>0$.  Then equality \eqref{eq:cRform} shows that $X_{2M}R$ is Hermitian.  Therefore $B$ is Hermitian and $R_{22}=\ol Y$.  Set $F=cR$ and use Lemma 2 in~\cite{bradler2018graph}.
\end{proof}

On behalf of all authors, the corresponding author states that there is no conflict of interest.

\end{document}